\documentclass[twocolumn,prb,aps,psfig,showpacs,preprintnumbers,superscriptaddress]{revtex4}
\usepackage{graphicx}
\usepackage{epstopdf}
\usepackage{float}
\usepackage{color}
\usepackage{amsmath}

\begin{document}
\title{Experimental evidence of a collinear antiferromagnetic ordering in the frustrated CoAl$_2$O$_4$ spinel}

\author{B. Roy}
\affiliation{Ames Laboratory and Department of Physics and Astronomy, Iowa State University, Ames, Iowa 50011, USA}
\author{Abhishek Pandey}
\affiliation{Ames Laboratory and Department of Physics and Astronomy, Iowa State University, Ames, Iowa 50011, USA}
\author{Q. Zhang}
\affiliation{Ames Laboratory and Department of Physics and Astronomy, Iowa State University, Ames, Iowa 50011, USA}
\author{T. W. Heitmann}
\affiliation{The Missouri Research Reactor Center, University of Missouri, Columbia, Missouri 65211, USA}
\author{D. Vaknin}
\affiliation{Ames Laboratory and Department of Physics and Astronomy, Iowa State University, Ames, Iowa 50011, USA}
\author{D. C. Johnston}
\affiliation{Ames Laboratory and Department of Physics and Astronomy, Iowa State University, Ames, Iowa 50011, USA}
\author{Y. Furukawa}
\affiliation{Ames Laboratory and Department of Physics and Astronomy, Iowa State University, Ames, Iowa 50011, USA}

\date{\today}

\begin{abstract} 
    Nuclear magnetic resonance (NMR), neutron diffraction (ND), x-ray diffraction, magnetic susceptibility $\chi$ and specific heat measurements on the frustrated $A$-site spinel CoAl$_2$O$_4$ compound reveal a collinear antiferromagnetic ordering below $T_{\rm N}$ = 9.8(2)~K\@. 
   A high quality powder sample characterized by  x-ray diffraction that indicates a relatively low Co-Al inversion parameter $x = 0.057(20)$ in (Co$_{1-x}$Al$_{x}$)[Al$_{2-x}$Co$_x$]O$_4$, shows a broad maximum around 15~K in $\chi(T)$ and a sharp peak at $T_{\rm N}$  in heat capacity $C_{\rm p}$. 
   The average ordered magnetic moment of Co$^{2+}$ ($S$ = 3/2) ions at the $A$-site is estimated to be 2.4(1)~$\mu_{\rm B}$ from NMR and 1.9(5)~$\mu_{\rm B}$ from ND which are smaller than the expected value of 3~$\mu_{\rm B}$ for $S = 3/2$ and $g$ = 2. 
   Antiferromagnetic spin fluctuations and correlations in the paramagnetic state are revealed from the $\chi$, NMR and ND measurements, which are due to spin frustration and site inversion effects in the system. 
   The ND data also show short-range dynamic magnetic ordering that persists to a temperature that is almost twice $T_{\rm N}$. 

\end{abstract}


\maketitle

\section{INTRODUCTION}

\begin{figure}[b]
  \includegraphics[width=2.5in]{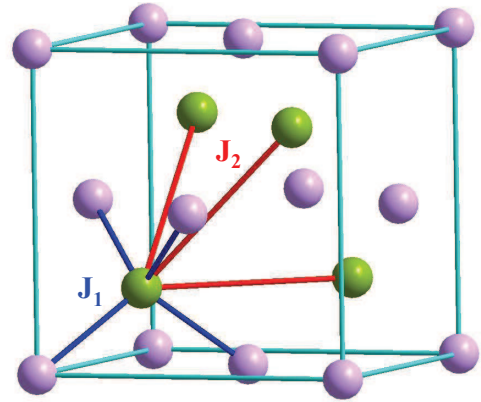}
	\caption{(Color online) Diamond lattice formed by Co atoms of cubic spinel CoAl$_2$O$_4$,  composed of two interpenetrating face-centered cubic (fcc) sublattices shown by purple and green spheres.
     The nearest-neighbor interaction  $J_1$ between the two fcc sublattices and  the next-nearest-neighbor interaction $J_2$ within the same fcc sublattice  are shown.} 
	\label{fig:Figure_A-site_Structure}
\end{figure}

   The investigation of magnetism in the insulating $A$-site spinel-structure compound CoAl$_2$O$_4$ has a long history which started with a suggestion of long-range antiferromagnetic (AFM) ordering below 4~K from magnetic susceptibility $\chi$ and neutron diffraction (ND) measurements performed by Roth.\cite{Roth1964a} 
    The magnetism of this system originates from Co$^{2+}$ ions [($e_{\rm g})^{\rm 4} $($t_{\rm {2g}})^{\rm 3}$] with spin $S$ = 3/2 at the tetrahedral $A$-sites which form a diamond lattice (Fig.~\ref{fig:Figure_A-site_Structure}).  
    The general formula for a normal oxide spinel compound is $A$[$B_2$]O$_4$ where the $A$ and $B$ atoms occupy the tetrahedral and octahedral holes between nearly cubic-close-packed O layers perpendicular to the [111] direction. 
     Later on, the interest shifted to the spin frustration effects in the $A$-site spinel compounds after the ground state of the diamond lattice AFM was investigated theoretically.\cite{Bergman2007}
    The diamond lattice is composed of two interpenetrating face-centered cubic (fcc) sublattices as shown in 
 Fig.~\ref{fig:Figure_A-site_Structure}. 
    The frustration is caused by the next-nearest-neighbor AFM interaction $J_2$ which couples nearest-neighbor sites of each fcc sublattice of the diamond structure, while the nearest-neighbor interaction  $J_1$ between the two fcc sublattices alone does not induce any frustration for AFM ordering (Fig.~\ref{fig:Figure_A-site_Structure}).   

    Bergman \emph{et al.}\cite{Bergman2007} pointed out theoretically that the ground state of the diamond lattice is a N\'eel-type antiferromagnet for a ratio of $J_{2}/J_{1}<1/8 = 0.125$. 
    The ground state changes to a spiral spin-liquid state for $J_{2}/J_{1}>1/8$.  $J_{1}$ and $J_{2}$ for CoAl$_2$O$_4$ are reported to be  $J_{1} = 0.92(1)$ meV $\approx$ 10.67~K and with $J_{2} = 0.101(2)$ meV  $\approx 1.17$~K, $J_{2}/ J_{1}$ = 0.11.\cite{Zaharko2011} 
    Since the ratio is close to the critical ratio of $J_{2}/J_{1} = 1/8$, CoAl$_2$O$_4$ is considered to be located in the critical region between the AFM and the spiral spin-liquid states. 

     Experimental observations of the ground state of CoAl$_2$O$_4$ are contradictory. 
     Tristan \emph{et al.}\cite{Tristan2005} reported from  $\chi$ and heat capacity $C_{\rm p}$ measurements on powder samples that a spin-glass state is realized below 4.8~K with a high frustration parameter\cite{Ramirez1994} $f = |\theta_{\rm CW}|/T_{\rm M}$ = 22, where  $\theta_{\rm CW}$ is the Curie-Weiss temperature in the  Curie-Weiss law and $T_{\rm M}$ is the spin-glass transition temperature. 
    On the other hand, Suzuki \emph{et al.} proposed a possible spin-liquid ground state below 9~K (Ref~\onlinecite{Suzuki2007}) with $f = |\theta_{\rm CW}|/T^* = 10$ where $T^*$ is defined as the temperature at which a broad peak of $C_{\rm p}$ is observed. 
     Zaharko \emph{et al.}\ suggested an unconventional magnetically ordered phase in which the spin-liquid correlations were observed  from neutron powder diffraction\cite{Zaharko2010} (below 5~K) and single-crystal ND experiments\cite{Zaharko2011} (below 8~K).
    MacDougall \emph{et al.}\ also carried out a ND experiment on a single crystal of CoAl$_2$O$_4$ which showed a sharp cusp at 6~K in the $T$-dependence of $\chi$ and an AFM state below 6.5~K.\cite{MacDougall2011}     
    They observed strong diffusive scattering in the AFM state where the phase transition was
suggested to be first order nature.\cite{MacDougall2011}  
    The origin of this ambiguity is probably site inversion, as has been pointed out by Zaharko \emph{et al}.\cite{Zaharko2011} 
    The degree of site inversion between the $A$- and $B$-sites is usually defined by an inversion parameter $x$ in the formula $A_{1-x}B_x[B_{2-x}A_x$]O$_4$, which here corresponds to (Co$_{1-x}$Al$_x$)[Al$_{2-x}$Co$_x$]O$_4$. 
    The $x$ value depends delicately on the method of sample preparation. 
    The $x$ values of the polycrystalline samples reported by Suzuki \emph{et al.}\cite{Suzuki2007}, Tristan \emph{et al.}\cite{Tristan2005, Tristan2008} and Zaharko \emph{et al.}\cite{Zaharko2010} are 0.04(1), 0.08--0.104 and 0.17, respectively. 
    The $x$ values for the single-crystal samples studied by Zaharko\cite{Zaharko2011} and  MacDougall\cite{MacDougall2011} are 0.08 and 0.02(4), respectively. 

    Quite recently Hanashima \emph{et al.}\cite{Hanashima2013} carried out a systematic study of the effects of site inversion on the magnetic properties of (Co$_{1-x}$Al$_x$)[Al$_{2-x}$Co$_x$]O$_4$ where  $x$ is controlled from 0.0467 to 0.153 by changing the heat treatment conditions. 
    The $\chi$($T$)  of the sample with the lowest $x$ value $x = 0.0467$ shows a broad maximum at 14~K and does not show any signature of magnetic ordering down to 2~K, similar to that reported by Suzuki \emph{et al.},\cite{Suzuki2007} suggesting a spin-liquid state. 
     With increasing $x$, the broad maximum moves to lower temperature (10~K for $x = 0.0643$), and for $x > 0.101$, a clear cusp in the $T$-dependence of $\chi$ is observed which is attributed to a spin-glass transition below $T_{\rm g} \approx 4.5$~K. 
    In the intermediate region 0.0791 $< x <$ 0.847, a coexistence of the spin-liquid state and spin-glass state is proposed. 
    Hanashima \emph{et al.} conclude that with increasing $x$ the spin-liquid state collapses and the spin-glass order emerges for $x > 0.101$. 

    It is important to understand whether the intrinsic ground state of CoAl$_2$O$_4$ for small values of $x$ is a spin-glass state, an antiferromagnet or a spin-liquid state due to the spin frustration. 
    To investigate the ground state of CoAl$_2$O$_4$ one needs a high-quality sample with a small value of the inversion parameter $x$. 
    In this paper, we report measurements of crystallography, magnetization, $\chi$, specific heat, nuclear magnetic resonance (NMR) and ND on a well-characterized polycrystalline CoAl$_2$O$_4$ sample with inversion parameter $x = 0.057(20)$.    
    We conclude that CoAl$_2$O$_4$ exhibits collinear AFM ordering below the $T_{\rm N}$ = 9.8(2)~K\@, with an ordered moment of 2.4(1)~$\mu_{\rm B}$ from NMR and 1.9(5)~$\mu_{\rm B}$ from ND measurements.

\section{EXPERIMENTAL DETAILS}

    A polycrystalline sample of CoAl$_2$O$_4$ was synthesized by the solid state reaction method using ${\rm Co_3O_4}$ (99.9985\%) and ${\rm Al_2O_3}$ (99.995\%) from Alfa Aesar as the starting materials. As a precautionary measure to check the phase-purity of the starting materials, room temperature x-ray diffraction (XRD) measurement was performed on the ${\rm Co_3O_4}$ starting material.  
    The results revealed that the material was not single-phase and with a sizable presence of binary CoO. 
    The weight fraction of CoO obtained by two-phase Rietveld refinement of the powder XRD pattern was 1.4(1)\%.
     The molar fraction of CoO in ${\rm Co_3O_4}$ was considered while calculating the stoichiometry of cobalt in the starting mixture. 
     Stoichiometric mixtures of ${\rm Co_3O_4}$ and ${\rm Al_2O_3}$ were thoroughly mixed and pressed into a pellet. 
    The pressed pellet was placed inside an alumina crucible and heated to 875~$^{\circ}$C in 20~h, held there for 40~h and then cooled to 500~$^{\circ}$C in 100~h and held there for 80~h, and finally the sample was cooled by switching off the furnace. 
    The material was then thoroughly reground and repressed into a pellet. 
    The pellet was placed in the same alumina crucible and heated to 1000~$^{\circ}$C in 20~h, held there for 50~h, then cooled to 700~$^{\circ}$C in 60~h, held there for 2~h and finally slowly cooled to 400~$^{\circ}$C in 100~h and held there for 100~h. 
    At this time the furnace was switched off and the sample cooled to room temperature. The sample was persian blue in color.

     Structural characterization of the sample was performed at room temperature using powder XRD data obtained from a Rigaku Geigerflex powder x-ray diffractometer utilizing CuK$_{\alpha}$ radiation. 
     The {\tt FullPROF} package\cite{Carvajal1993} was used in the Rietveld XRD data refinement.
     Temperature- and applied magnetic field $H$-dependent magnetic measurements were performed using a magnetic properties measurement system (MPMS, Quantum Design, Inc.). 
   Heat capacity $C_{\rm p}$ was measured by the thermal relaxation method using a physical properties measurement system (PPMS, Quantum Design, Inc.).
 
    NMR measurements were conducted by probing the $^{27}$Al nuclei ($I$ = 5/2, $\gamma$/2$\pi$ = 11.09375~MHz/T, $Q$ = 0.149~barns) and the $^{59}$Co nuclei ($I$ = 7/2, $\gamma$/2$\pi$ = 10.03~MHz/T, $Q$ = 0.4~barns) by using a phase-coherent spin-echo pulse spectrometer.  
    The NMR spectrum was taken by recording the spin-echo signal intensity while sweeping the external magnetic field at a fixed resonance frequency. 
    The Co NMR spectrum under zero magnetic field was obtained by plotting the spin-echo intensity point by point at different fixed frequencies.

    Neutron diffraction measurements were performed using the triple-axis spectrometer (Triax) at the  University of Missouri Research Reactor (MURR). 
    Approximately 2 grams of CoAl$_2$O$_4$ powder was loaded into a vanadium sample cell and mounted on the cold head of an Advanced Research Systems closed-cycle refrigerator on Triax.  
    Powder ND patterns were obtained on the spectrometer operated with a Cu(220) monochromator at a wavelength of 1.348~{\AA}, a PG(002) analyzer set at the elastic position, and collimation of $60^\prime-40^\prime-40^\prime-80^\prime$.  
    Sapphire and Si filters were used upstream from the monochromator and none after the sample.  
    Additional scans with greater statistics were also obtained at the key positions in wavevector (${\bf q}$) space corresponding to the putative spiral or collinear magnetic orderings.  
The 2~g sample held in the same holder was also measured on the MURR high resolution neutron powder diffractometer (PSD), which utilizes a double-focusing perfect single Si monochromator at the (511) reflection to illuminate the sample with 1.4805~{\AA} wavelength neutrons. 
    Diffracted neutrons are detected by five linear position sensitive detectors. 
    The sample temperature was adjusted with a Leybold closed-cycle refrigerator with a 18 K base temperature. 

\section{CRYSTALLOGRAPHY}

\begin{figure}[tb]
  \includegraphics[width=3in]{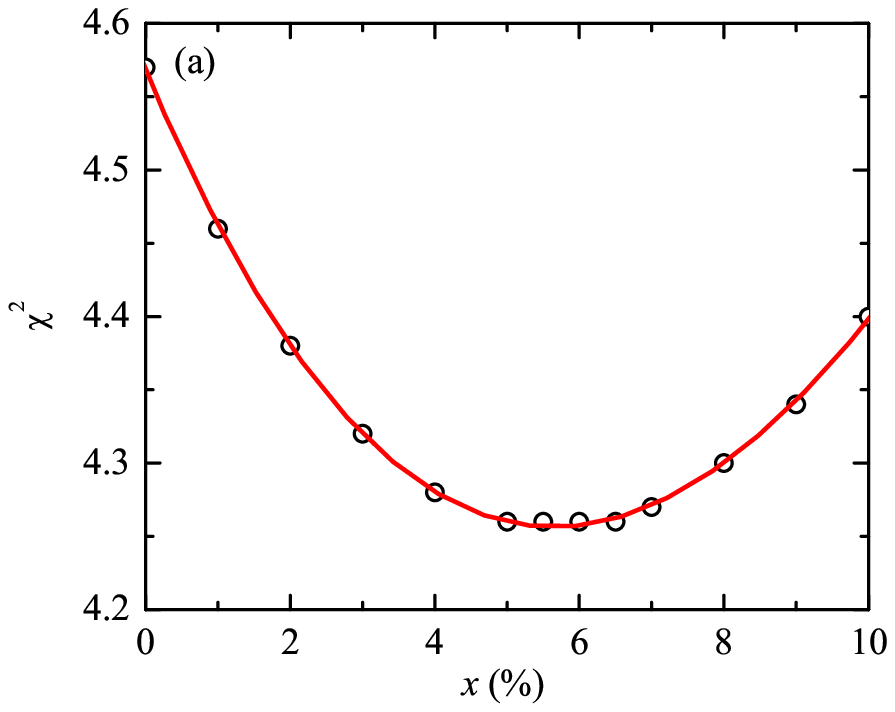}
	\includegraphics[width=3in]{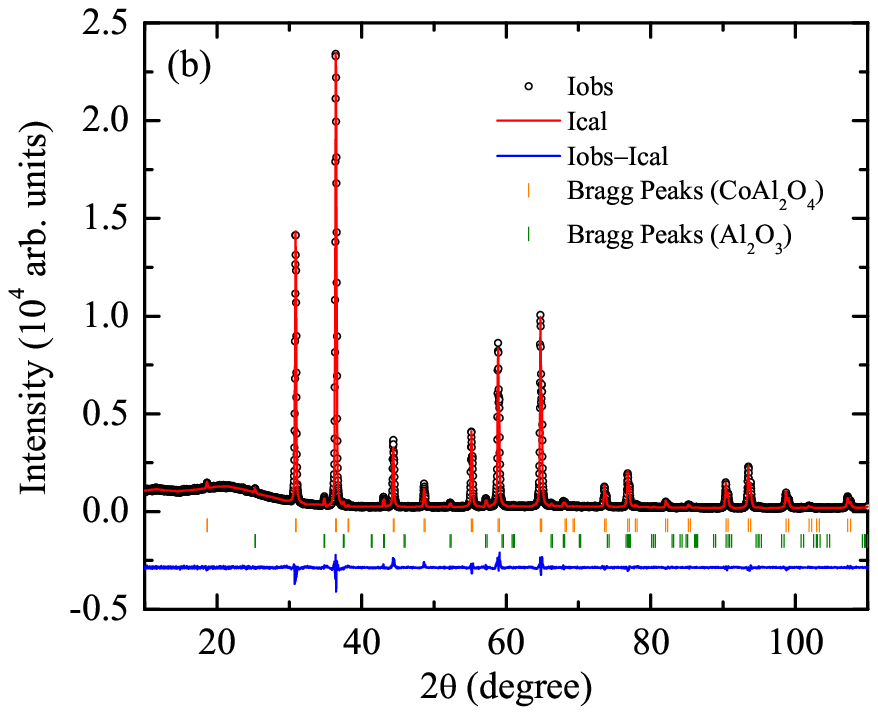}
	 \caption{(Color online) (a) Goodness of fit parameter $\chi^2$ versus site inversion parameter $x$ obtained by single-phase Rietveld refinement of our polycrystalline CoAl$_2$O$_4$ sample. 
       The red line is shown as eye guides. 
         (b) Room-temperature powder x-ray diffraction data, two-phase Rietveld refinement, Bragg positions and difference profile for CoAl$_2$O$_4$ and the Al$_2$O$_3$ impurity phase.} 
	\label{fig:Figure_XRD}
\end{figure}

    As described in the Sec. I, it was shown that the magnetic properties of CoAl$_2$O$_4$ depend delicately on the amount of site inversion present in the sample.\cite{Hanashima2013} 
    The goal in this work is to investigate the magnetic properties of a high-quality polycrystalline sample of CoAl$_2$O$_4$ having minimal site inversion. 
    To achieve this we synthesized a polycrystalline sample of CoAl$_2$O$_4$ using the controlled thermal treatments described in the previous section. 
    However, we could not achieve a completely site-inversion-free sample. 
    Rietveld analysis of the room-temperature powder XRD data of our sample showed a minor amount of site inversion corresponding to the formula (Co$_{1-x}$Al$_x$)[Al$_{2-x}$Co$_x$]O$_4$.     
    The goodness of fit parameter $\chi^2$ obtained by single-phase Rietveld refinements of the room-temperature powder XRD data versus site inversion $x$ is shown in Fig.~\ref{fig:Figure_XRD}(a). 
     From this figure we estimate $x = 0.057(20)$. 
    This value of $x$ is slightly smaller than that reported by Tristan \emph{et al.}\ (0.080--0.104) in Refs.~\onlinecite{Tristan2005} and \onlinecite{Tristan2008}  and Zaharko \emph{et al.}\ (0.08) in Ref.~\onlinecite{Zaharko2011},  and is equal , within error bars, to that reported by Suzuki \emph{et al.}\ (0.04) in Ref.~\onlinecite{Suzuki2007} and Hanashima \emph{et al.}\ (0.0467).\cite{Hanashima2013}

\begin{table}
\caption{Crystal data, Wyckoff positions and refined structural parameters along with the Rietveld refinement quality parameters for CoAl$_2$O$_4$.} 
\label{Table:Structure}
 \begin{ruledtabular}
		\begin{tabular}{l l }
		 Structure     & Spinel (${\rm MgAl_2O_4}$-type)  \\
		 Space group   & $Fd\bar{3}m$  (No. 227)\\
		 Lattice parameter & 8.1011(2)~\AA  \\
		 CoAl$_2$O$_4$ weight fraction & 96.5(2)\% \\ 
		 Inversion parameter ($x$) & 0.057(20) \\
		 \vspace{0.05cm}\\
		 Atom & Wyckoff Positions \\
		 \hline
		 Co & 8$a$ (1/8, 1/8, 1/8) \\
		 Al & 16$d$ (1/2, 1/2, 1/2) \\
		 O &  32$e$ ($x_{\rm O}, x_{\rm O}, x_{\rm O}$) \\
		   & where $x_{\rm O}$ = 0.2630(1)
		 \vspace{0.05cm}\\
		 Refinement Quality Parameters \\
		 \hline
		 $\chi^{2}$   & 2.49 \\
		 $R_{\rm p}$ (\%)  & 4.69  \\
		 $R_{\rm wp}$ (\%) & 6.37   \\
	\end{tabular}
\end{ruledtabular}
\end{table}

    Additionally, powder XRD data showed the presence of a minor secondary phase Al$_2$O$_3$ in our sample. 
    After estimating the site inversion parameter by single-phase Rietveld refinement we performed a two-phase Rietveld refinement with a fixed value of $x = 0.057$. 
    Figure~\ref{fig:Figure_XRD}(b) shows the results of the two-phase Rietveld refinement. 
    We achieved an excellent fit to the data, and the calculated weight fraction of Al$_2$O$_3$ was 3.5(2)\%. 
    The crystal and Rietveld refinement quality parameters are listed in Table~\ref{Table:Structure}.

\section{Magnetization and magnetic susceptibility}

\begin{figure}[t]
  \includegraphics[width=3in]{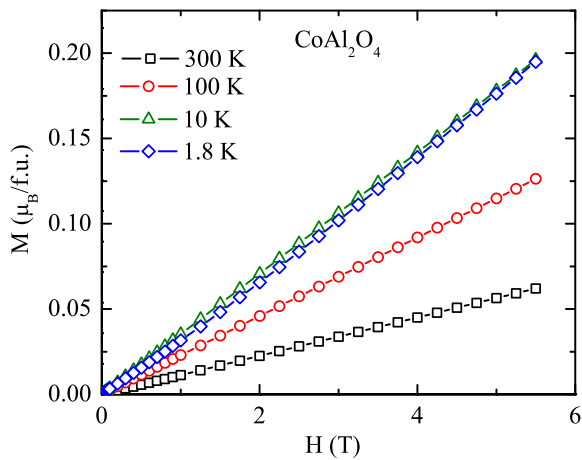}
	 \caption{(Color online) Isothermal magnetization $M$ versus applied magnetic field $H$ for CoAl$_2$O$_4$ measured at four different temperatures between 1.8 and 300~K.} 
	\label{fig:Figure_MH}
\end{figure}

   Isothermal magnetization $M$ versus $H$ data of our polycrystalline CoAl$_2$O$_4$ sample at four different temperatures are shown in Fig.~\ref{fig:Figure_MH}. 
   The $M(H)$ data at $T = 10, 100$ and 300~K are proportional to $H$ in the entire field range of measurement. 
    However, the $M(H)$ data at 1.8~K shows nonlinear behavior with positive curvature in the entire field range of measurement. 
   This behavior is typical of Heisenberg AFMs with anisotropy as observed from the ND measurements.\cite{Zaharko2011}  

\begin{figure}[t]
  \includegraphics[width=3in]{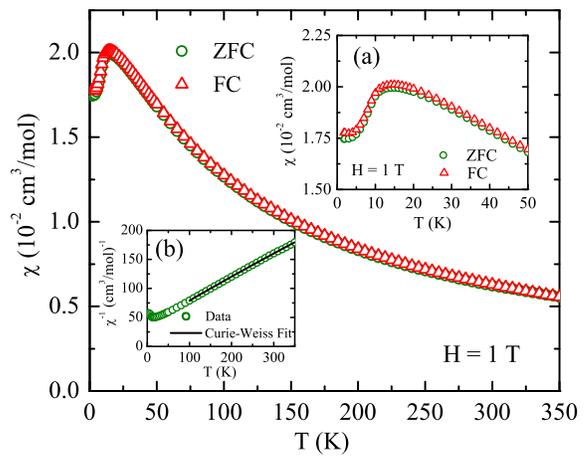}
	 \caption{(Color online) Zero-field-cooled (ZFC) and field-cooled (FC) magnetic susceptibility $\chi$ versus temperature $T$ for CoAl$_2$O$_4$. 
    The ZFC and FC data nearly overlap each other in the entire $T$-range of measurement. 
    Inset (a): Expanded plot of $\chi(T)$ below 50~K. 
    Inset (b): Inverse magnetic susceptibility $\chi^{-1}$ versus $T$. Solid line is Curie-Weiss fit to the data as described in the text.} 
	\label{fig:Figure_Susceptibility}
\end{figure}

    The $\chi(T)$ of our sample measured under zero-field-cooled (ZFC) and field-cooled (FC) conditions in the presence of an applied magnetic field $H = 1$~T is shown in Fig.~\ref{fig:Figure_Susceptibility}. 
   The $\chi(T)$ exhibits a broad maximum centered at $T$ = 15(1)~K suggesting the presence of AFM correlations in the material. 
    The ZFC and FC $\chi(T)$ data nearly overlap in the entire temperature range of measurement [Inset (a), Fig.~\ref{fig:Figure_Susceptibility}], which is consistent with the result on polycrystalline CoAl$_2$O$_4$ reported by Suzuki \emph{et al.}\ with a low inversion parameter $x = 0.04(2)$.\cite{Suzuki2007} 
    Our $\chi(T)$ data are also consistent with the results for the $x = 0.0467$ sample of Hanashima \emph{et al.},\cite{Hanashima2013} except below 9.6~K where there is a rather significant bifurcation of ZFC and FC data in their sample. 
    As discussed in Sec.~I, it has been shown in previous reports that the low-temperature magnetic properties of CoAl$_2$O$_4$ are very sensitive to the amount of site inversion between the $A$-site and the $B$-site in the samples. 
    Samples with relatively large inversion parameters ($x \ge 0.08$) have been reported to show spin-glass-like behaviors,\cite{Tristan2005} while samples with $x \le 0.05$ have been argued to have spin-liquid ground states.\cite{Suzuki2007, Hanashima2013}

\begin{table*}
\caption{Parameters obtained from magnetic susceptibility $\chi(T)$ data and their fit using high temperature Curie-Weiss law described in the text for polycrystalline CoAl$_2$O$_4$. 
   Data for some other CoAl$_2$O$_4$ samples reported in the literature have been included for comparison.  
   The listed parameters in the table are site inversion parameter $x$, magnetic ordering temperature $T_{\rm M}$, characteristic temperature for spin-liquid state  $T^*$,  Curie-Weiss temperature $\theta_{\rm CW}$, frustration parameter $f = |\theta_{\rm CW}/T_{\rm M}|$ or $|\theta_{\rm CW}/T^*|$, Curie constant $C$, effective paramagnetic moment $\mu_{\rm eff}$ obtained from $C$ assuming $g$ = 2 and effective spectroscopic splitting factor $g$ obtained assuming $S=3/2$ for the Co$^{2+}$ ions.} 
\label{Table:Magnetism}
 \begin{ruledtabular}
		\begin{tabular}{l c c c c c c c c l}
		 $x$ &  $T_{\rm M}$ & $T^*$ & $\theta_{\rm CW}$ & $f $ & $C$        & $\mu_{\rm eff}$ & $g$  & Ground  & Ref.\\
		 (\%)& (K)      & (K)   & (K)          &                            & (cm$^3$K/mol)&  ($\mu_{\rm B}$/f.u.)&      &State&                \\
		 \hline
		      &   & &  $-$73(8)      & 8(1)    & 2.1(1)     & 4.1(1)          & 2.12(5)   &   & This work\footnotemark[1]\\ 
		 \raisebox{1.5ex}{5.7(2.0)}        &  \raisebox{1.5ex}{9.8(2)}          & &  $-$104(9)           & 11(1)                      & 2.52(6)        & 4.49(5)          & 2.32(3)   &  \raisebox{1.5ex} {AFM}  & This work\footnotemark[2]\\ [2ex] 

		 4(2) &  &  9\footnotemark[3]        & $-$89(6)           & 9.9(7)                     & 2.5(1)        & 4.45(8)         & 2.30(4)    &Spin Liquid     & [\onlinecite{Suzuki2007}] \\
		 4.67 & & 9.6\footnotemark[4]         & $-$93.0            & 9.7                        & 2.38          & 4.36            & 2.25          & Spin Liquid & [\onlinecite{Hanashima2013}] \\
		 2(4) & 6.5  &        & $-$109(1)	         & 18.0(2)                    & 2.99(4)       & 4.89(3)         & 2.53(2)    & AFM   & [\onlinecite{MacDougall2011}] \\ 
		 8(3) &4.8(2)  &     & $-$104(2)          & 22(1)                      & 2.7(1)        &4.65(9)           & 2.40(4)   & Spin Glass   & [\onlinecite{Tristan2005}] \\
		 8 &8    &  &  $-$94(1)          & 12(1)                      & 2.7(1)        &4.63(2)           & 2.40(1)    & Unconventional    & [\onlinecite{Zaharko2011}, \onlinecite{Maljuk2009}] \\
	\end{tabular}
\footnotetext[1]{Estimated with $\chi_0$ = $6(2)\times 10^{-4}$~cm$^3$/mol. }
\footnotetext[2]{Obtained assuming $\chi_0$ = 0. }
\footnotetext[3]{Determined from the peak position in the heat capacity. }
\footnotetext[4]{Defined by the temperature at which ZFC and FC susceptibilities deviates from each other.  }
	\end{ruledtabular}
\end{table*}

    The $\chi(T)$ data between 100 and 350~K were fitted by the Curie-Weiss law [inset~(b) in Fig.~\ref{fig:Figure_Susceptibility}], 
\begin{equation}
\chi(T) = \frac{C}{T-\theta_{\rm CW}} + \chi_{0},
\end{equation} 
where $C$ is the Curie constant and $\chi_0$ is the $T$-independent orbital contribution to $\chi$. 
    The fitted value of $\chi_{0}$ is $6(2)\times 10^{-4}$~cm$^3$/mol, which is in good agreement with Van Vleck susceptibility $4(2)\times 10^{-4}$~cm$^3$/mol for Co$^{2+}$ ions at the $A$-site estimated from NMR measurements as will be described in Sec. VI. B. 
    This value is almost one order of magnitude greater than the diamagnetic susceptibility $\chi_{\rm dia}$ = $ -7.4\times 10^{-5}$~cm$^3$/mol estimated from the reference compound ${\rm ZnAl_2O_4}$ (Ref. \onlinecite{Tristan2005}).  
     The fitted values of $C$, $\theta_{\rm CW}$ and the effective moment $\mu_{\rm eff}$ and $g$-factors determined  
from $C$ are listed in Table~\ref{Table:Magnetism}. 
    For comparison,  parameters for some other CoAl$_2$O$_4$ samples reported in the literature are also included in Table~\ref{Table:Magnetism}. 
    The $C$ obtained for our sample is larger than the value 1.875~cm$^3$K/mol expected for spin $S = 3/2$ and $g = 2$. 
    However, our value of $C$ is smaller than those reported in the literature for this material (Table~\ref{Table:Magnetism}). 
    If we consider that magnetism in CoAl$_2$O$_4$ is entirely due to Co$^{2+}$ spins with $S = 3/2$ then the calculated value of powder-averaged $g$-factor is 2.12(5) which is again considerably smaller than that reported for CoAl$_2$O$_4$ in the literature (Table~\ref{Table:Magnetism}) and $g$ = 2.26 reported from electron spin resonance (ESR) measurements.\cite{Hagiwara2010}
    It should be noted, however, that C value strongly depends on $\chi_0$. 
    If one assumes $\chi_0$ = 0, then $C$ = 2.52(6)~cm$^3$K/mol was obtained with $\theta_{\rm CW}$ = $-104(9)$~K (Table~\ref{Table:Magnetism}). 
Using this value of $C$, the $g$-factor is estimated to be 2.32(2) which is slightly larger than  $g$ = 2.26 obtained from the ESR measurements.\cite{Hagiwara2010}    
    It appears that the value of the frustration parameter $f$ depends sensitively on the degree of antisite disorder present in CoAl$_2$O$_4$ samples. 
    The $f$ value of our sample is lower than those reported in the literature (Table~\ref{Table:Magnetism}).

\section{Heat capacity}

\begin{figure}[t]
  \includegraphics[width=3in]{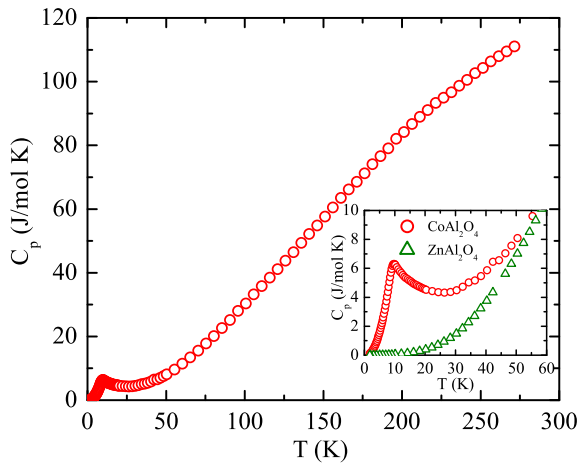}
	 \caption{(Color online) Heat capacity $C_{\rm p}$ versus temperature $T$ for CoAl$_2$O$_4$. 
   Inset: The $C_{\rm p}(T)$ of CoAl$_2$O$_4$ at an expanded scale near magnetic transition temperature $T_{\rm N}$ along with $C_{\rm p}(T)$ of nonmagnetic analog compound ${\rm ZnAl_2O_4}$ (Ref.~\onlinecite{Tristan2008}). 
   The $C_{\rm p} (T)$ data of ${\rm ZnAl_2O_4}$ provided by Tristan \emph{et al.} were for $T \geq 10$~K. We extrapolated the data below 10~K using $C_{\rm p} = \beta~T^3$.
}
	\label{fig:Figure_HC}
\end{figure}

\begin{figure}[t]
  \includegraphics[width=3in]{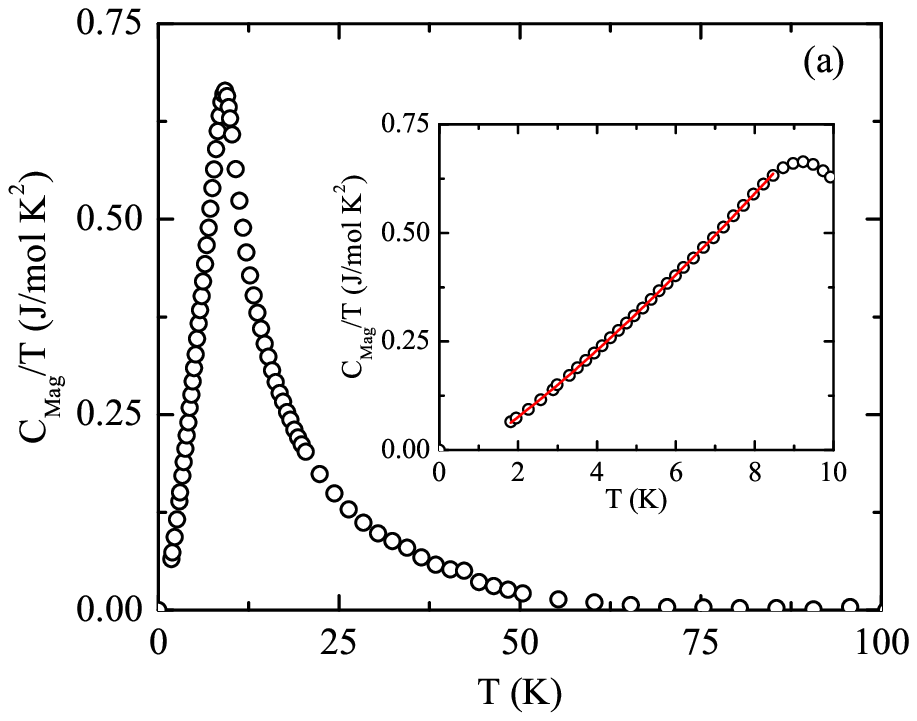}
  \includegraphics[width=3in]{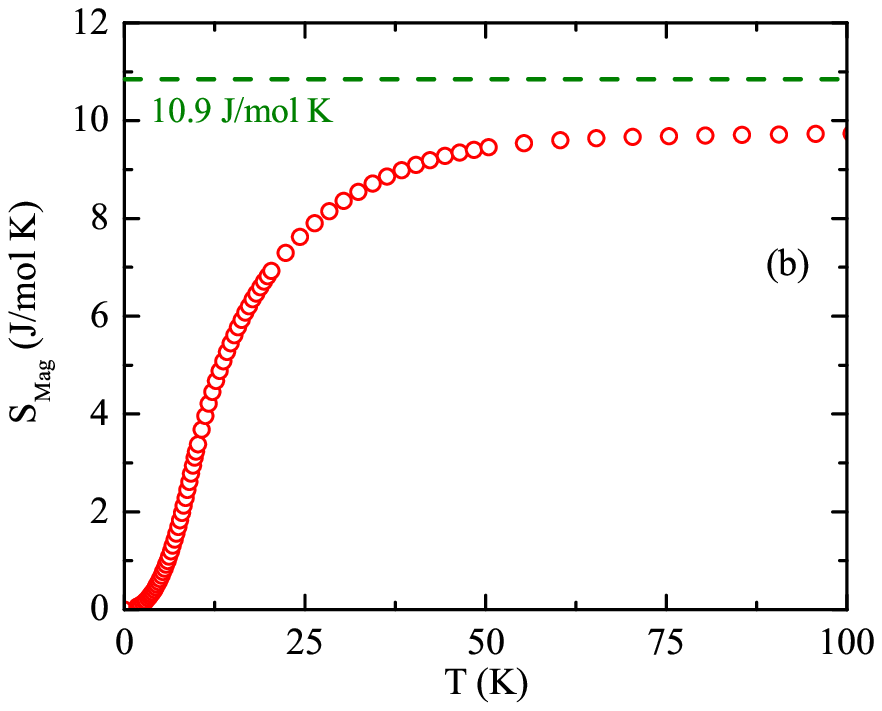}
	 \caption{(a) (Color online) Magnetic contribution to the heat capacity $C_{\rm Mag}$ over temperature $T$ versus $T$. An additional data point $C_{\rm Mag}/T = 0$ at $T = 0$ was inserted for better estimation of the magnetic entropy. 
Inset : $C_{\rm Mag}/T$ versus $T$ below 10~K. Solid line shows $C_{\rm Mag}/T$ $\propto$ $T^{1.23}$ dependence. 
(b) Magnetic entropy $S_{\rm Mag}$ versus $T$. 
   The horizontal dashed line represents the calculated high-$T$ limit $S_{\rm Mag}$ = 10.9~J/mol K\@.} 
	\label{fig:Figure_HC-Mag}
\end{figure}

    The $C_{\rm p}$ versus $T$ for CoAl$_2$O$_4$ is shown in Fig.~\ref{fig:Figure_HC}. 
    Low-temperature $C_{\rm p}(T)$ data of CoAl$_2$O$_4$  along with those of the nonmagnetic reference compound ${\rm ZnAl_2O_4}$ (Ref. \onlinecite{Tristan2008}) are shown in inset on an expanded scale. 
    The $C_{\rm p}(T)$ data of CoAl$_2$O$_4$ show a sharp peak centered at $T$ = 9.7(2)~K suggesting a magnetic transition at this temperature. 
   The magnetic contribution $C_{\rm Mag}(T)$ to the total heat capacity was obtained by subtracting the $C_{\rm p}(T)$ data of ${\rm ZnAl_2O_4}$ from that of CoAl$_2$O$_4$ [Fig.~\ref{fig:Figure_HC-Mag}(a)]. 

     The magnetic contribution $S_{\rm Mag}(T)$ to the total entropy was calculated from $C_{\rm Mag}$ using 
\begin{equation}
S_{\rm Mag}(T) = \int_{0}^{T}{\frac{C_{\rm Mag}(T)}{T}}dT.
\end{equation}  
The $T$-dependence of $S_{\rm Mag}$ is shown in Fig.~\ref{fig:Figure_HC-Mag}(b), where we have also indicated the expected high temperature limit
\begin{eqnarray}
S_{\rm Mag} (T \rightarrow \infty) 
& = & (1-x)R~{\rm ln}(2S+1)   \nonumber \\
& = & 0.943R~{\rm ln}4  = 10.9~{\rm J/mol~K}.  
\end{eqnarray}
where we consider only the contribution from the $A$-site Co$^{2+}$ ions with $S=3/2$ for our sample with $x$ = 0.057 for simplicity.
     From the $S_{\rm Mag}(T)$ data presented in the Fig.~\ref{fig:Figure_HC-Mag}(b) we find that 
   (i) $S_{\rm Mag}$ does not reach the expected saturation limit and its value at 100~K is 9.74~J/mol~K which is $\approx 11\%$ smaller than 10.9 J/mol~K\@. 

   (ii) The value of $S_{\rm Mag}$ at $T_{\rm N} = 9.7$~K is 3.11~J/mol~K which is only $\approx 32\%$ of it's high-$T$ value. 
    This clearly shows that a major part of the magnetic entropy is released between 10 and 50~K, which indicates the persistence of strong dynamic short-range magnetic order above $T_{\rm N}$ up to approximately 50~K\@.  
 
    There are contradictory reports in the literature about the $T$-dependence of $C_{\rm p}$ below the magnetic ordering temperature. 
    While Tristan \emph{et al.}\cite{Tristan2005} and MacDougall \emph{et al.}\cite{MacDougall2011} reported that $C_{\rm p}$ varies as $T^2$, Suzuki \emph{et al.}\ find that this dependence is $T^{2.5}$.\cite{Suzuki2007} 

    In our sample we find that $C_{\rm Mag}$ below $T_{\rm N}$ follows $C_{\rm Mag}$ $\propto$ $T^{2.23}$ which is approximately mid way between  the $T^2$ and $T^{2.5}$ power law dependences. 
     Our observation is inconsistent with the linear $T$-dependence of $C_{\rm p}$ expected for spin glass systems\cite{Meschede1980, Walker1977} and the exponent is slightly lower than the $T^{7/3}$ dependence predicted by Bergman \emph{et al.}\ for spiral spin-liquid frustrated diamond-lattice antiferromagnets.\cite{Bergman2007}

\section{Nuclear magnetic resonance}
\subsection{$^{27}$Al-NMR}
\begin{figure}[t]
  \includegraphics[width=3.7in]{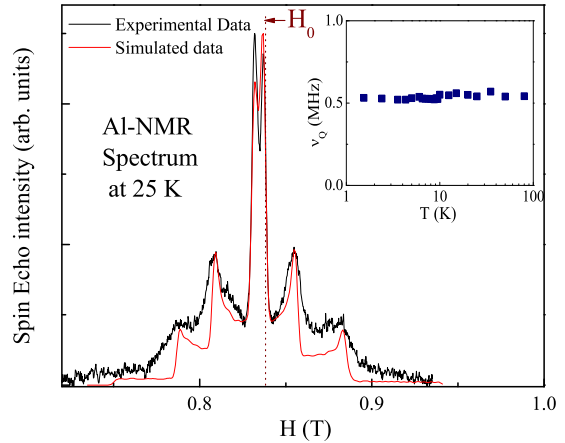}
     \caption{(Color online) Field-swept $^{27}$Al-NMR spectrum at 25~K and  $\nu_{0}$ = 9.3 MHz.  
    The red line shows the simulated spectrum with $\nu_{\rm Q}$ = 0.55 MHz and $\eta$ = 0. 
    The inset shows the temperature $T$ independence of the quadrupole frequency $\nu_{\rm Q}$.} 
              \label{fig:1Al-NMR-specfit}
\end{figure}

    Figure~\ref{fig:1Al-NMR-specfit} shows the field-swept $^{27}$Al-NMR spectrum at $T$ = 25~K and resonance frequency $\nu_0$ = 9.3~MHz. 
    Since our sample consists of grains with randomly oriented crystal axes, the spectrum is a powder pattern. 
     The spectrum can be explained by a combination of a large Zeeman interaction ${\cal H}_{\rm Z}$ and a small quadrupole interaction ${\cal H}_{\rm Q}$\@. 
     The nuclear spin Hamiltonian can therefore be expressed as 
\begin{equation}
              {\cal H} = {\cal H}_{\rm Z} + {\cal H}_{\rm Q},
                \label{eq:Hamiltonian}
\end{equation}
where 
\begin{math}
{\cal H}_{\rm Z} = - \gamma \hbar {H}_{0} (1+K){I}_{z}   
\end{math}
and 
\begin{math}
{\cal H}_{\rm Q} = \frac{e^2qQ}{4I(2I-1)} [(3I^{2}_{z} - I^2)+\frac{1}{2}\eta(I^{2}_{+} + I^{2}_{-})], 
\end{math}
$Q$ is the quadrupole moment of the $^{27}$Al nucleus and $K$ is the NMR shift which is a sum of isotropic ($K_{\rm iso}$) and axial ($K_{\rm ax}$) parts. $\eta$ is the asymmetry parameter for the electric field gradient (EFG) defined by $\frac{ \frac{\partial^2 V}{\partial X^2}- \frac{\partial^2 V}{\partial Y^2}}{ \frac{\partial^2 V}{\partial Z^2}}$, $|{ \frac{\partial^2 V}{\partial Z^2}}| > | { \frac{\partial^2 V}{\partial Y^2}}| > | { \frac{\partial^2 V}{\partial X^2}} |$, $eq$ = $|{ \frac{\partial^2 V}{\partial Z^2}}|$ and the quadrupole frequency $\nu_{\rm Q}$ = $\frac{3e^2qQ}{2I(2I-1)h}$. 
   The $z$ and $Z$ axes are the quantization axes for the Zeeman and quadrupole interactions, respectively. 
    The observed spectrum is well reproduced by a simulated spectrum shown by a red line in Fig.~\ref{fig:1Al-NMR-specfit} with $\nu_Q$ = 0.55 MHz and $\eta$ = 0. 
    The value of $\nu_Q$ is independent of $T$ in the measured range of 1.8 to 300~K within our experimental uncertainty as shown in the inset of Fig.~\ref{fig:1Al-NMR-specfit}. 
    In contrast, $K_{\rm iso}$ and $K_{\rm ax}$ show weak temperature dependences.  
   The weak $T$-dependences of the NMR shifts are consistent with previous data reported by Miyatani \emph{et al.}\cite{Miyatani1965} who measured the $^{27}$Al-NMR spectrum in CoAl$_2$O$_4$ with the continuous wave NMR method above 77 K, although the $K_{\rm ax}$ component was not detected in the previous measurement.

\begin{figure}[t]
  \includegraphics[width=3.7in]{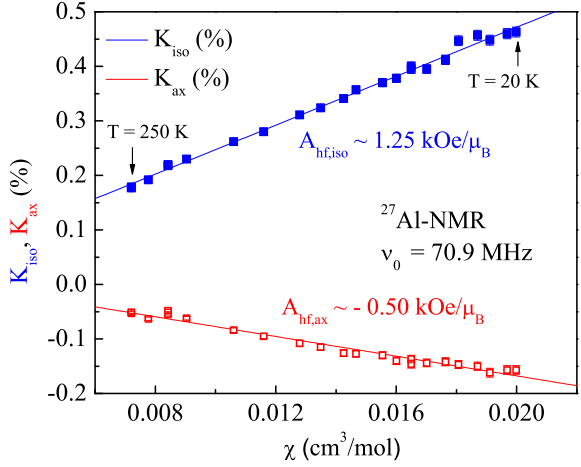} 
        \caption{(Color online) $^{27}$Al-NMR shifts ($K_{\rm {iso}}$ and $K_{\rm {ax}}$) versus $\chi$ with $T$ as an implicit parameter.  The solid lines show the linear fit using the Eq.~(\ref{eq:K-Chi}).}
          \label{fig:2Al-NMR-K}
\end{figure}

     The $T$-dependences of $K_{\rm iso}$ and $K_{\rm ax}$ follow the $T$-dependence of $\chi$ as shown in Fig.~\ref{fig:2Al-NMR-K} where the NMR shifts are plotted as a function of $\chi$ with $T$ as an implicit parameter. 
   Here the NMR shifts are determined by measuring the NMR spectrum with a higher resonance frequency of 70.9 MHz to improve the accuracy. 
    The NMR shift has contributions from the $T$-dependent spin part  $K_{\rm spin}$ and a small $T$-independent orbital part  $K_{\rm 0}$. $K_{\rm spin}$($T$) is related to the spin susceptibility with hyperfine coupling constant $A_{\rm{hf}}$ as follows:
\begin{equation}
    K(T) = K_{\rm{0}} + \frac{A_{\rm{hf}}}{N_{\rm{A}}} \chi(T),
     \label{eq:K-Chi}
\end{equation}
where $N_{\rm{A}}$ is the Avogadro number. 
The isotropic ($A_{\rm{hf,iso}}$) and axial ($A_{\rm{hf,ax}}$) parts of the hyperfine coupling constants for the Al nucleus are evaluated from the slope of the $K$-$\chi$ plots (Fig.~\ref{fig:2Al-NMR-K}) to be 1.25(2) kOe/$\mu_{\rm{B}}$ and $-0.50(2)$ kOe/$\mu_{\rm{B}}$ respectively, and 
$K_{\rm{0}}$ is 0.023(4)\% and 0.013(3)\% for the isoisotropic and axial parts, respectively.
   Classical dipolar field calculations indicate that the hyperfine field cannot be due to the classical dipolar field at the Al-sites from neighboring ${\rm{Co}}^{2+}$  ($S$ = 3/2) spins, but is rather due to the transferred hyperfine field due to finite spin transfer from ${\rm{Co}}^{2+}$ spins. 
   The anisotropy in the hyperfine field suggests that anisotropic orbitals such as the 2$p$ and 3$p$ orbitals on Al atoms are polarized by the spin transfer from the neighboring Co atoms. 
 
\begin{figure}[t]
  \includegraphics[width=3.7in]{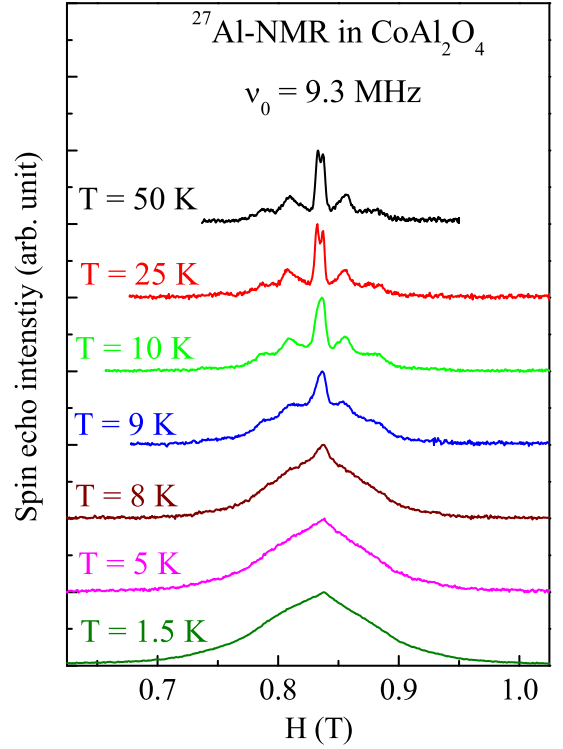} 
     \caption{(Color online) Field-swept $^{27}$Al-NMR spectra at various temperatures with resonance frequency $\nu_0$ = 9.3 MHz. } 
        \label{fig:3Al-NMR-specT}
\end{figure}

\begin{figure}[t]
  \includegraphics[width=3.7in]{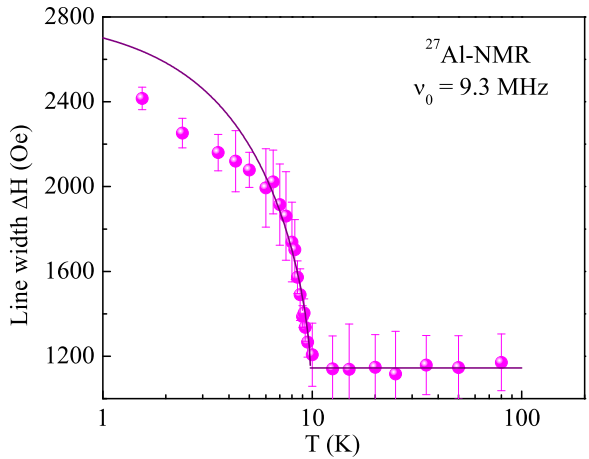} 
        \caption{(Color online) $T$-dependence of the line width at 10\% of the peak intensity. 
The solid purple line shows the fit by Eq.~(\ref{eq:OrderParameter}) using data in the temperature range of 7.5 K $< T <$ 10 K.} 
        \label{fig:4Al-NMR-linewidth}
\end{figure}

    The spectrum starts to broaden suddenly below $\approx$ 10~K, as shown in Fig.~\ref{fig:3Al-NMR-specT}. 
   The broadening indicates a static magnetic ordering of the ${\rm{Co}}^{2+}$ spins which produce a finite static internal field at the Al-sites. 
    The symmetric line broadening indicates an AFM state. 
    Figure~\ref{fig:4Al-NMR-linewidth} shows the $T$ dependence of the line width ($\Delta H$) determined at 10 \% of the peak intensity. 
    Since the broadening of the $^{27}$Al-NMR line width in our polycrystalline sample below $\approx$ 10~K is produced by the Co$^{2+}$ spin moments in the magnetically ordered state, the $T$ dependence of the line width reflects that of the sublattice magnetization in the AFM state. 
     The N\'eel temperature $T_{\rm{N}}$ is determined by fitting $\Delta H$ near $T_{\rm{N}}$ in Fig~\ref{fig:4Al-NMR-linewidth} by 
\begin{equation}
         \Delta H (T) = \Delta H_0 \left(1-\frac{T}{T_N}\right)^\beta + c,
         \label{eq:OrderParameter}
\end{equation}
where  $\Delta H_0$ $\approx$ 1669 Oe and $c$ $\approx$ 1145 Oe is a constant determined by averaging the nearly $T$-independent $\Delta H$ above $T_{\rm{N}}$. 
    By fitting data in the temperature range of 7.5 K $< T <$ 10 K, by Eq.~(\ref{eq:OrderParameter}), $T_{\rm{N}}$ is estimated to be  9.8(2) K and $\beta$ = 0.65(5). 
 
      In general, one can drive the universality class/spin dimensionality from the critical exponent $\beta$. 
      The $\beta$ = 0.65(5) is much larger than $\beta$ = 0.367 for the 3 dimensional (D) Heisenberg model, 0.345 for the 3D XY model, and 0.326 for the 3D Ising model.\cite{de -Jongh}
      It should be noted that our $\beta$ value strongly depends on the fitting range. 
      By decreasing the minimum temperature of the fit, the $\beta$ was found to decrease with a minimum value  $\beta$ = 0.48(5) obtained by using all data below 10~K. 
       The $\beta$ value is also estimated to be 0.255(44)  from ND data for 3~K $<$ $T$ $<$ 10~K as will be described in Sec. VII. 
      Since  the $\beta$ value strongly depends on the fitting range, we cannot discuss the details of the universality class or spin dimensionality in CoAl$_2$O$_4$ based on our NMR and ND data.

\subsection{$^{59}$Co-NMR}

\begin{figure}[t]
  \includegraphics[width=3.7in]{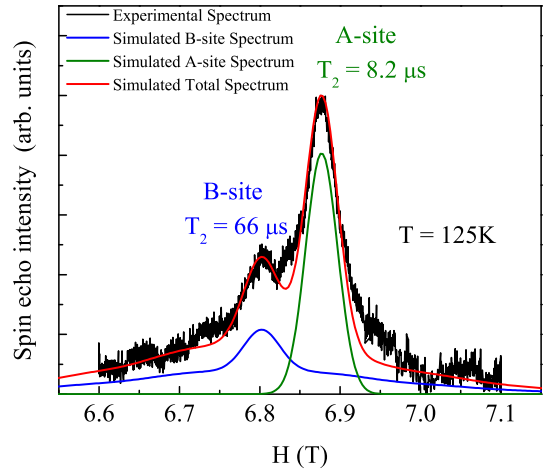} 
        \caption{(Color online) Field-swept $^{59}$Co-NMR spectrum at $T$ = 125~K and resonance frequency $\nu_0$ =  70.2~MHz. The two peaks correspond to the Co ions occupying the $A$-site (magnetic ${\rm Co^{2+}}$ ion) and the $B$-site (nonmagnetic ${\rm Co^{3+}}$ ion), respectively. 
The transverse relaxation time $T_{\rm 2}$ = 8.2~$\mu$s and $T_{\rm 2}$ = 66~$\mu$s for the $A$- and $B$-sites, respectively.} 
         \label{fig:5Co-NMR-specfit}
\end{figure}

     In the paramagnetic state at $T$ = 125~K, we observe two peaks in the $^{59}$Co-NMR spectrum as shown in Fig.~\ref{fig:5Co-NMR-specfit}. 
   With decreasing $T$, the peak observed at $H~\approx$ 6.9 T shifts to higher magnetic field and the other peak at $H~\approx$ 6.8~T moves to slightly lower magnetic field. 
   While the former peak disappeared below 10~K, the latter peak could be observed down to the lowest temperature of 1.5~K.
   These behaviors are very similar to $^{59}$Co-NMR data for the magnetic ${\rm Co^{2+}}$ ($S = \frac{3}{2}$) at the $A$-site and nonmagnetic low-spin ${\rm Co^{3+}}$ [($t_{\rm {2g}})^{\rm 6} $($e_{\rm g})^{\rm 0}$, $S = 0$]  at the $B$-site in the spinel-structure antiferromagnet ${\rm Co_3O_4}$ with $T_{\rm{N}}$ = 34 K\@.~\cite{Fukai1996} 
    Thus the former signal, which vanishes due to very large internal field at the ${\rm Co^{2+}}$-site, can be assigned to be the Co at the tetrahedral $A$-site and the latter signal can be attributed to the Co occupying the octahedral $B$-site.

    From the intensities of the Co signals occupying the $A$- and the $B$-sites, corrected by their respective longitudinal and transverse relaxation times ($T_{\rm 1}$ and $T_{\rm 2}$, respectively), we have evaluated the inversion parameter to be $x \approx$ 0.06(2). 
    This value is in good agreement with that from the x-ray analysis.  
    We notice that the $T$ dependence and the magnitude of the NMR shift of the Co signal from the $B$-sites in CoAl$_2$O$_4$ is very close to that observed in the Co NMR signal from the $B$-sites in  ${\rm Co_3O_4}$ where the ${\rm Co^{3+}}$ ions are in a nonmagnetic state with low spin $S$ = 0.\cite{Fukai1996} 
    From these results, we may consider that  the Co ions occupying the $B$-site in CoAl$_2$O$_4$ could be ${\rm Co^{3+}}$ with low spin state as in the case of Co ions at the $B$-site in ${\rm Co_3O_4}$. 
    This could be possible if  we assume that the deviation from oxygen stoichiometry exists as in (Co$_{1-x}$Al$_{x}$)[Al$_{2-x}$Co$_x$]O$_{4+\delta}$ to satisfy charge compensation of the compound. 
    However, we cannot rule out completely that the observed nonmagnetic Co is from an impurity phase, although it seems unlikely.  
      
\begin{figure}[t]
  \includegraphics[width=3.5in]{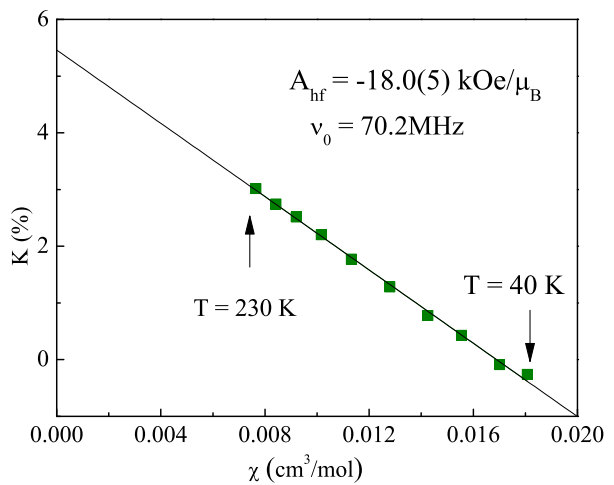} 
            \caption{(Color online) $^{59}$Co-NMR shift $K$ versus $\chi$ with $T$ as an implicit parameter. 
      The solid line is a linear fit.} 
              \label{fig:6Co-NMR-K}
\end{figure}

     The hyperfine coupling constant of Co at the $A$-site is estimated from the $K$-$\chi$ plot in Fig.~\ref{fig:6Co-NMR-K} to be $-$18.0(5) kOe/$\mu_{\rm B}$  which is considerably smaller than the value of $-$105 kOe/$\mu_{\rm B}$  for a free ${\rm Co^{2+}}$ ion. 
    Such a small Co hyperfine coupling constant at the $A$-site has been reported previously to be $-$18.3~kOe/$\mu_{\rm B}$  and $-$15.7~kOe/$\mu_{\rm B}$  in the antiferromagnetic ${\rm Co_3O_4}$  ($T_{\rm N}$ = 34~K) (Ref.~\onlinecite{Fukai1996}) and in ferrimagnetic ${\rm CoCr_2O_4}$ ($T_{\rm C}$ = 97~K) (Ref.~\onlinecite{Tsuda1968}), respectively. 
    The small coupling constant was well explained by taking the second-order orbital moment due to the spin-orbit interaction into consideration. 
   Here we present a similar discussion as in the previous papers.\cite{Tsuda1968, Fukai1996}
    
     The major contributions to the magnetic hyperfine coupling constant at the $^{59}$Co nucleus are dominated by the $d$ core polarization ($A_{\rm hf}^{\rm core}$), dipolar ($A_{\rm hf}^{\rm dip}$) and orbital ($A_{\rm hf}^{\rm orb}$) terms:\cite{Tsuda1968, Fukai1996, Furukawa1999}
\begin{equation}
       A_{\rm hf} = A_{\rm hf}^{\rm core} + A_{\rm hf}^{\rm dip} + A_{\rm hf}^{\rm orb} .
      \label{eq:TotalHyperfineCoupling}
\end{equation}
Due to the crystal field effect, the ground state of the ${\rm Co^{2+}}$($d^7$) ion occupying the tetrahedral $A$-site has four electrons in the two lower  $e_{\rm g}$ orbitals and three in the upper three $t_{\rm 2g}$ orbitals. 
     This electronic configuration results in $A_{\rm hf}^{\rm dip}$ = 0.~\cite{Fukai1996} 
     For a free ${\rm Co^{2+}}$ ion, the $d$ core polarization coupling constant is $-$105~kOe/$\mu_{\rm B}$.~\cite{Abragam1955} 
     Due to quenching of the orbital angular momentum $L_z$, there is no finite contribution of the $L_z$ in the first order. 
     But if we consider the second order perturbation of the spin-orbit interaction, the orbital quenching is partially lifted. 
     This contribution due to the spin-orbit coupling has a positive sign and is estimated to be 85.1~kOe/$\mu_{\rm B}$ for ${\rm Co^{2+}}$ ions at the tetrahedral sites.\cite{Fukai1996}  
     From Eq.~(\ref{eq:TotalHyperfineCoupling}), we obtain $A_{\rm hf}$ = $-$19.9~kOe/$\mu_{\rm B}$, which can reasonably explain the experimental value of $-$18.0(5)~kOe/$\mu_{\rm B}$. 

     The temperature independent orbital part of the NMR shift for the Co$^{2+}$ at the A-site is estimated to be $K_{\rm orb}$ = 5.46(3)\% from the $K$-$\chi$ analysis. 
    The $K_{\rm orb}$ relates to Van Vleck susceptibility $\chi_{\rm VV}$ for Co$^{2+}$ at the A-site as $K_{\rm orb}$ = $\frac{A_{\rm orb}}{N_{\rm A}}$$\chi_{\rm VV}$ where  $A_{\rm orb}$ is the orbital hyperfine coupling constant given by \cite{Narath_Hyper}
 \begin{equation}
       A_{\rm orb} = 2 \mu_{\rm B} \langle \frac{1}{r^3} \rangle .
      \label{eq:OrbitalHyperfineCoupling}
\end{equation}
Here $\langle \frac{1}{r^3} \rangle $ is an average of $\frac{1}{r^3}$ over 3$d$ electrons. 
   Using  $\langle \frac{1}{r^3} \rangle $ = 6.02 a.u. for Co$^{2+}$ (Ref. \onlinecite{Freeman1965}), $\chi_{\rm VV}$ is estimated to be $4(2)\times 10^{-4}$ cm$^3$/mol which is in good agreement with $T$ independent magnetic susceptibility discussed in Sec. IV.

\begin{figure}[t]
  \includegraphics[width=3.5in]{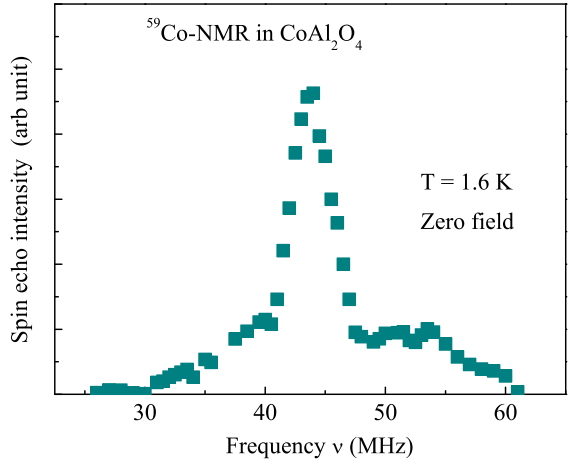} 
         \caption{(Color online)Zero-field $^{59}$Co-NMR spectrum under zero external field at 1.6~K\@. } 
           \label{fig:7Co-NMR-zerofield}
\end{figure}

   Below $T_{\rm N}$, we observe an NMR signal of Co nuclei at the $A$-site under zero magnetic field, which again confirms long-range magnetic ordering.  
   Figure~\ref{fig:7Co-NMR-zerofield} shows the zero-field $^{59}$Co-NMR spectrum at 1.6~K\@.
    A similar zero-field $^{59}$Co-NMR spectrum at the $A$-site has been previously reported for the spinel compound  ${\rm Co_3O_4}$ in the AFM state.\cite{Fukai1996} 
   The resonance frequency $\nu_0$ = 43.5(4)~MHz at the peak intensity point is lower than $\nu_0$ = 55.7~MHz in ${\rm Co_3O_4}$.\cite{Fukai1996} 
    The internal field $H_{\rm int}$ at the Co $A$-site is evaluated to be 43.3(4) kOe at 1.6~K\@. 
With the value of $A_{\rm hf}$ = $-$18.0~kOe/$\mu_{\rm B}$ deduced from the $K$-$\chi$ plot, 
the ordered moment $<\mu>$ of ${\rm Co^{2+}}$ is estimated to be 2.4(1)~$\mu_{\rm B}$/Co using the relation $H_{\rm int}$ = $|A_{\rm hf}|$$<\mu>$. 
    This value is smaller than the expected value of 3~$\mu_{\rm B}$/Co for $S$ = 3/2 of ${\rm Co^{2+}}$ ions with $g$ = 2 and is larger than the value of 1.58~$\mu_{\rm B}$/Co reported from the ND study done previously by Zaharko \emph{et al.}\cite{Zaharko2011} 
    The shoulder-like structures of the spectrum are due to a distribution of ${\rm Co^{2+}}$ spin moments in the AFM state.

\subsection{$^{27}$Al spin lattice relaxation rates}
\begin{figure}[t]
  \includegraphics[width=3.5in]{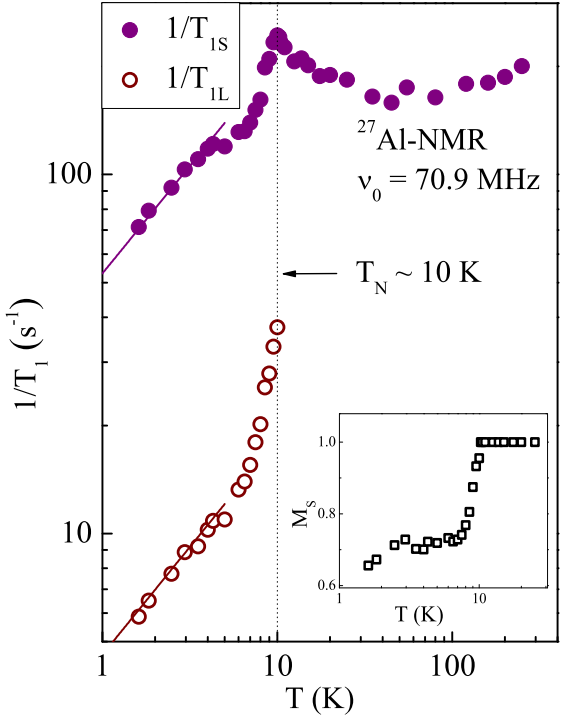} 
             \caption{(Color online)  $T$ dependence of the short and long nuclear spin-lattice relaxation rates, $1/T_{\rm {1S}}$ and $1/T_{\rm {1L}}$, respectively. 
Below 4 K, both $1/T_{\rm {1S}}$ and $1/T_{\rm {1L}}$ show $T^{0.6}$ power law behavior shown by the solid lines. The inset shows the $T$ dependence of $M_{\rm S}$ which corresponds to the fraction of  $1/T_{\rm {1S}}$ defined in Eq. (\ref{eq:2ComponentsforT1forI3/2}).} 
               \label{fig:8Al-NMR-T1}
\end{figure}

In order to investigate the dynamical behavior of the ${\rm Co^{2+}}$ spins in CoAl$_2$O$_4$, $^{27}$Al spin-lattice relaxation rates ($1/T_{\rm 1}$) were measured at different temperatures. In the case of $^{27}$Al ($I$ = 5/2), the nuclear magnetization recovery of the central transition after a $\pi$/2 saturation pulse is given by \cite{Narath1967}
\begin{equation}
\begin{split}
\frac{M(\infty)-M(t)}{M(\infty)} &= (0.029 e^ {-t/T_{1}}) +(0.18 e^ {-6t/T_{1}})\\
                                             & + (0.79 e^ {-15t/T_{1}})
\end{split}
\label{eq:T1forI3/2}
\end{equation}
where $M(t)$ is the nuclear magnetization at time $t$ after the saturation pulse. 
   The experimental recovery curves above $T_{\rm N}$ were well fitted by the equation, while below $T_{\rm N}$ the equation could not fit the experimental data due to distributions of $T_{\rm 1}$. 
    Then we tentatively assumed that the experimental recovery curve is composed of two components, one with a short relaxation time $T_{\rm{1S}}$ and another with long relaxation time $T_{\rm{1L}}$ and we carried out a fit using the following equation:
\begin{equation}
\begin{split}
\frac{M(\infty)-M(t)}{M(\infty)} &= M_{\rm S}[(0.029 e^ {-t/T_{\rm{1S}}}) +(0.18 e^ {-6t/T_{\rm{1S}}})\\
                                             & + (0.79 e^ {-15t/T_{\rm{1S}}})]+M_{\rm L}[(0.029 e^ {-t/T_{\rm{1L}}})\\
                                            & +(0.18 e^ {-6t/T_{\rm{1L}}}) +(0.79 e^ {-15t/T_{\rm{1L}}})]
\end{split}
\label{eq:2ComponentsforT1forI3/2}
\end{equation}
where $M_{\rm S}$ + $M_{\rm L}$ = 1. The temperature dependence of $M_{\rm S}$ is shown in the inset of Fig.~\ref{fig:8Al-NMR-T1}. 

    Figure~\ref{fig:8Al-NMR-T1} shows the temperature dependence of $1/T_{\rm{1S}}$ and $1/T_{\rm{1L}}$. 
    With decreasing $T$, $1/T_{\rm{1S}}$ decreases and shows a minimum around 50 K and exhibits a sharp peak at $T$ = 10.0(5)~K. 
    The sharp peak of $1/T_{\rm 1}$ is due to critical slowing down of spin fluctuations expected for a second order phase transition, which again confirms the AFM long-range magnetic ordering at $T_{\rm{N}}$. 
    Below 10 K, $1/T_{\rm 1}$ rapidly falls and then decreases slowly where both $1/T_{\rm{1S}}$ and 
$1/T_{\rm{1L}}$ show $T^{\rm 0.6}$ power-law behavior (shown by the straight lines in the figure) below 4~K. 
    In the antiferromagnetically ordered state, $1/T_{\rm 1}$ is mainly driven by scattering of magnons, leading to $T^{\rm 3}$ or $T^{\rm 5}$ power-law temperature dependences due to a two- or three-magnon Raman process, respectively. 
    The weak $T$ dependence of $1/T_{\rm 1}$ $\sim$ $T^{\rm 0.6}$ below 4~K cannot be explained by the magnon-scattering, and suggests the presence  of other magnetic fluctuations in the magnetically ordered state, which could originate from spin frustrations and/or short-range magnetic correlations revealed by the neutron scattering measurements described below.

\begin{figure}[t]
  \includegraphics[width=3.5in]{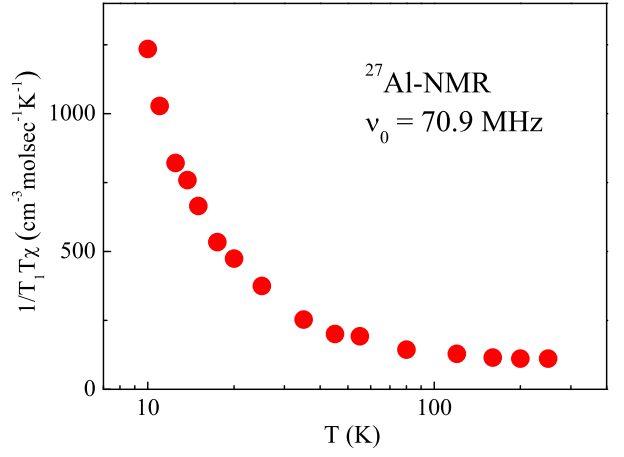} 
             \caption{(Color online) $1/T_{\rm 1}T$$\chi$ versus temperature $T$ for a resonance frequency of 70.9 MHz. $1/T_{\rm 1}T$$\chi$ increases with decreasing temperature indicating the growth of AFM spin correlations.} 
               \label{fig:9Al-NMR-T1TChi}
\end{figure}

In order to see spin fluctuation effects in the paramagnetic state, it is useful to re-plot the data by changing the vertical axis from $1/T_{\rm 1}$ to $1/T_{\rm 1}T\chi$ as shown in Fig.~\ref{fig:9Al-NMR-T1TChi}. $1/T_{\rm 1}T$ can be expressed in terms of the imaginary part of the dynamic susceptibility $\chi^{\prime\prime}(\vec{q}, \omega_0)$ per mole of electronic spins as:\cite{Moriya1963, Mahajan1998}
\begin{equation}
\frac{1}{T_1T}=\frac{2\gamma^{2}_{N}k_{\rm B}}{N_{\rm A}^{2}}\sum_{\vec{q}}|A(\vec{q})|^2\frac{\chi^{\prime\prime}(\vec{q}, \omega_0)}{\omega_0}, 
\label{eq:T1TChi}
\end{equation}
where the sum is over the wave vectors $\vec{q}$ within the first Brillouin zone, $A(\vec{q})$ is the form factor of the hyperfine interactions and $\chi^{\prime\prime}(\vec{q}, \omega_0)$  is the imaginary part of the dynamic susceptibility at the Larmor frequency $\omega_0$.  
    On the other hand,  the uniform $\chi$ corresponds to the real component 
 $\chi^{\prime}(\vec{q}, \omega_0)$ with $q$ = 0 and $\omega_0$ = 0. 
    Thus a plot of $1/T_{\rm 1}T\chi$ versus $T$ shows the $T$ dependence of  $\sum_{\vec{q}}|A(\vec{q})|^2\chi^{\prime\prime}(\vec{q}, \omega_0)$ with respect to that of the uniform susceptibility $\chi^{\prime}$(0, 0). 
    For high temperatures above 100 K, $1/T_{\rm 1}T\chi$ is a nearly constant showing that the temperature dependence of $\sum_{\vec{q}}|A(\vec{q})|^2\chi^{\prime\prime}(\vec{q}, \omega_0)$ scales to that of $\chi^{\prime}$(0, 0).  
   On the other hand, with decrease in temperature, $1/T_{\rm 1}T\chi$ starts to increase below 100 K which is $\sim$~10 times higher than $T_{\rm N}$. 
   The increase of the $1/T_{\rm 1}T\chi$ at $T>>T_{\rm N}$ cannot be simply attributed to the critical slowing down of spin fluctuations near $T_{\rm N}$. 
    This implies $\sum_{\vec{q}}|A(\vec{q})|^2\chi^{\prime\prime}(\vec{q}, \omega_0)$ increases  more than $\chi^{\prime}$(0, 0), which is due to a growth of spin fluctuations with $q$ $\neq$ 0, most likely with AFM wave vector $q$ = $Q_{\rm AF}$, even at $T$ much higher than $T_{\rm N}$. 
    Thus we conclude that strong AFM spin fluctuations are realized in a wide temperature region up to $\sim$100~K in the paramagnetic state.    

\section{Neutron Diffraction}

\begin{figure}[t]
  \includegraphics[width=3.0in]{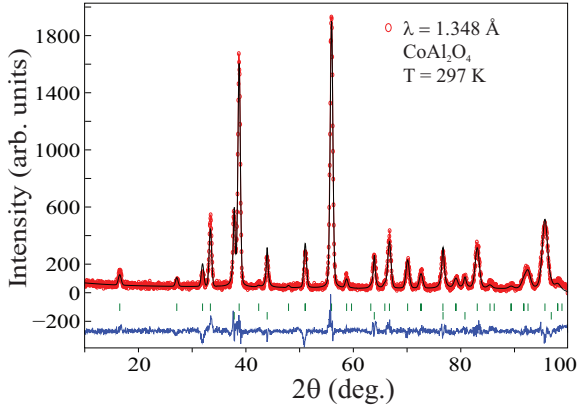} 
\caption{(Color online) Neutron diffraction intensity versus scattering angle  $2\theta$ at room temperature (red circles) and the calculated refined pattern (solid black line). 
The vertical green tick marks indicate the peak positions of Bragg reflections and the lower blue line shows the difference between the model calculations and measured data. } 
\label{ND-Rietveld}
\end{figure}

   Analysis of a powder neutron diffraction pattern taken at room temperature (Fig.\ \ref{ND-Rietveld}) is consistent with a single $Fd\bar{3}m$ phase of CoAl$_2$O$_4$.  
    Due to the relatively small difference in the neutron coherent scattering lengths between Al and Co, the Rietveld refinement of the powder diffraction pattern is not sensitive to the Co-Al site inversion discussed in section III. 
 
\begin{figure}[t]
  \includegraphics[width=3.0in]{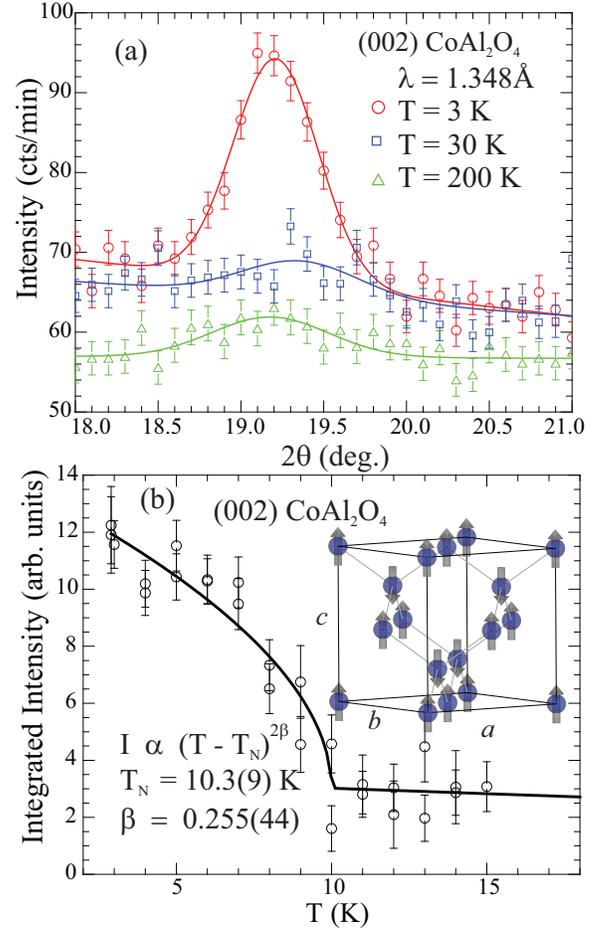} 
\caption{(Color online) (a) Intensity versus scattering angle ($2\theta$) of the (002) magnetic reflection at various temperatures demonstrating the long-range magnetic order.  
    Temperature dependence of background in the  $2\theta$ range shown in (a) indicates possible short-range magnetic order above $T_{\rm N}$.  
     The weak peak above the transition is nonmagnetic in origin (most likely due to a strong nuclear (004) reflection by $\lambda/2$ contamination in the beam) and persists to room temperature.
     (b) Integrated intensity of the (002) as a function of temperature. 
    The solid line is a fit to a power-law with $T_{\rm N}=10.3(9)$~K and $\beta = 0.255(44)$.  
Inset: Antiferromagnetic structure below $T_{\rm N}$. 
Magnetic and crystallographic unit cell are same with $a = b = c$. 
     } 
\label{fig:NRD}
\end{figure}

   A prominent (002) reflection, not allowed by the chemical structure, emerges  below 10~K as shown in Fig.~\ref{fig:NRD}(a) and its intensity increases upon cooling to the base temperature of 3~K. 
    Measurable intensity increases are also observed for the (111), (113), (222), and (133) nuclear Bragg reflections between scans taken at 15 and 3~K (not shown). 
    Whereas the (002) peak is slightly broader than the Bragg peaks in its vicinity, the (111) magnetic peak is much broader than the (111) nuclear peak in agreement with similar observations of single crystals.\cite{MacDougall2011,Zaharko2011} 
    The unusual broadening along the (111) reflection, (reported for a single crystal \cite{Zaharko2011}), makes the observation of the (222) reflection virtually impossible with the small amount of sample we used in this study.  
    The emergence of the (002) and the increase in intensities of the Bragg reflections, that fall off systematically with the increase in $2\theta$ (due to magnetic form factor), are consistent with long-range AFM ordering of the Co spins at the $A$-sites with a magnetic space group $F\bar{4}3m$. 
    Based on the temperature dependence of the integrated intensity of the (002)  (see Fig.~\ref{fig:NRD}(b)),  $T_{\rm N}$ is estimated to be 10.3(9) K, which is in agreement with the value estimated from NMR and specific heat measurements within experimental uncertainty.
    The diamond structure of Co$^{2+}$  consists of two interpenetrating fcc sublattices, with origins at (0,0,0) (spin up) and at ($\frac{1}{4},\frac{1}{4},\frac{1}{4}$) (spin down), as shown in the inset of Fig.\ \ref{fig:NRD}(b). 
    The four spins in one sublattice are parallel to the [001] direction  and the other four in the other sublattice are antiparallel to those.    
    This AFM structure is similar to that observed for the $A$-site of Co$^{2+}$ in Co$_3$O$_4$, for which an average magnetic moment 3.26 $\mu_{\rm B}$/Co at 4.2 K was obtained from ND analysis.~\cite{Roth1964b}    
     Our detailed analysis of the magnetic diffraction pattern yields a 1.9(5)~$\mu_{\rm B}$/Co average ordered magnetic moment, lower than 3 $\mu_{\rm B}$/Co expected for Co$^{2+}$ ($S = 3/2$, $g$ = 2), in agreement with 2.4(1)~$\mu_{\rm B}$/Co from the above NMR measurements within errors.

\begin{figure}[t]
  \includegraphics[width=3.0in]{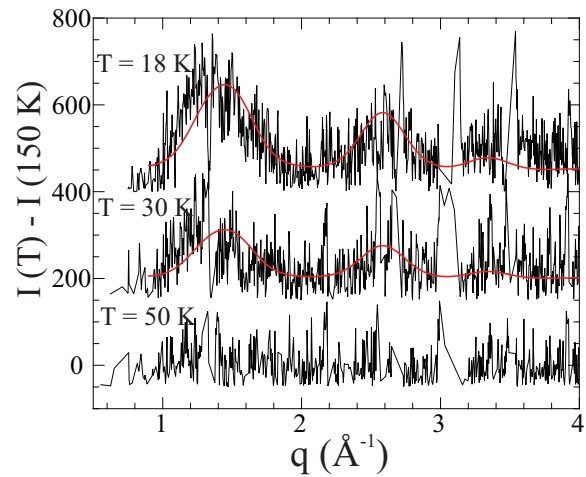} 
\caption{(Color online) Elastic neutron scattering intensity versus scattering wavevector $q$ at various temperatures above $T_{\rm N}$, as indicated, after subtraction of a similar scan at $T$ = 150~K\@. 
The broad peaks at 18~K and 30~K are evidence for short-range AFM order that persists up to about $ T= 50$~K\@. Sharp spikes in the subtracted patterns are due to temperature shifts of prominent nuclear Bragg reflections. Solid lines are calculated magnetic diffraction patterns assuming the underlying magnetic structure shown in Fig.\ \ref{fig:NRD}(b) with 6\% random Co$^{2+}$ vacancies. }
\label{SRO-Mag}
\end{figure}

    Figure\ \ref{SRO-Mag} shows the diffraction intensity versus momentum transfer [$q=4\pi\sin(\theta)/\lambda$] at three temperatures above $T_{\rm N}$ after subtracting a similar data set taken at higher $T$ = 150~K\@.
    The patterns indicate remnant dynamic short-range order (SRO) above the transition that disappears between 30 and 50~K\@. 
     The pattern, shown in Fig.\  \ref{SRO-Mag} at $T$ = 18~K, is very similar to that observed by Krimmel \emph{et al.}~\cite{Krimmel2006a} at $T$ = 1.6~K for a sample with  6--8\% site inversion which does not show long-range magnetic ordering down to 1.6~K\@. 
    In order to see the effects of the site inversion on the short-range magnetic ordering, 
we have calculated the $q$-dependence of the intensity by assuming 6\% random Co$^{2+}$ vacancies in a unit cell.
     In the calculation, we assumed the magnetic structure at the base temperature with the 6\% random vacancies, and allowed for uniform broadening of all magnetic peaks beyond the spectrometers resolution.
    The calculated results shown by solid lines in Fig. \ref{SRO-Mag} reasonably reproduce the observed patterns.  
   This suggests that  the SRO above $T_{\rm N}$ originates mainly from the site inversion effects. 
    It is known that the site inversion can be as high as 22\% (Ref. \onlinecite{Nakatsuka2003}) and such high inversion can eventually lead to a complete suppression of long-range magnetic ordering.~\cite{Krimmel2006a, Hanashima2013} 
   In this regard, it is interesting to note that the AFM transition temperature for Co$_3$O$_4$, associated with the ordering of the Co$^{2+} A$-site, is at $T_{\rm N}$ $\approx$ 30--40~K,~\cite{Roth1964a, Fukai1996, Suzuki2007, Roth1964b} which is very close to the temperature at which the SRO disappears in our sample. 
   In view of the site inversion playing a major role in determining $T_{\rm N}$, it is likely that reducing the inversion will enhance $T_{\rm N}$ in CoAl$_2$O$_4$.
    Of course, $T_{\rm N}$ will also be controlled by the spin frustration which has been pointed out to play an important role in the system. 
    To investigate the effect of spin frustration due to next-nearest neighbor exchange coupling ($J_2$) in detail, one needs an inversion-free CoAl$_2$O$_4$ sample,  although so far no one has obtained such a sample, with the lowest value of inversion parameter being at $x$ $\approx$ 0.02(4).\cite{MacDougall2011}
    Another candidate to study spin frustration effects in $A$-site spinels would be Co$_3$O$_4$ where the likelihood for Co$^{2+}$-Co$^{3+}$ inversion is negligible, with the $B$-site occupied by the nonmagnetic Co$^{3+}$.   
    However, since the ratio of the exchange coupling constants for Co$_3$O$_4$ is $J_2/J_1$  = 0.019 (with $J_1$ = 1.09~meV and $J_2$ = 0.02~meV) (Ref.~\onlinecite{Zaharko2011}) which is almost 6 times smaller than 0.109 for CoAl$_2$O$_4$ (Ref.~\onlinecite{Zaharko2011}), one expects much weaker spin frustration effects in Co$_3$O$_4$.

\section{Summary and conclusions}

    Our $C_{\rm p}(T)$, NMR and ND data clearly demonstrate that ${\rm CoAl}_2{\rm O}_4$ exhibits a collinear AFM ground state below $T_{\rm N}$ = 9.8(2)~K\@. 
   The $^{27}$Al-NMR spectrum exhibits a symmetric broadening below 10 K which indicates the presence of a long-range AFM ordered ground state for this compound. 
    Furthermore, the line width of the NMR spectrum which is proportional to the order parameter shows a continuous increase below 10~K\@, as expected for second order phase transition.  
    The presence of spontaneous magnetization below 10~K is confirmed directly by the observation of a $^{59}$Co NMR signal under zero magnetic field in the magnetically ordered state. 
    Using the internal field at the Co sites in the magnetically ordered state and the hyperfine coupling constant estimated from the $K$-$\chi$ analysis, an ordered moment at the $A$-site Co$^{2+}$ ion is estimated to be  2.4(1)~$\mu_{\rm B}$/Co.     
   The sharp peak of  $1/T_{\rm 1}$ at around 10~K further confirms a second-order phase transition to a long-range magnetically ordered state at the same temperature. 
   The results from the ND measurements are not only consistent with the NMR results but also indicate a collinear long-range AFM-ordered ground state for this compound with an ordered moment of 1.9(5)~$\mu_{\rm B}$/Co, consistent with the estimate from NMR.  

  The magnetic entropy $S_{\rm Mag}$ at $T_{\rm N}$ is considerably smaller than its value at high temperature, which indicates that strong AFM correlations occur above $T_{\rm N}$. 
    The emergence of AFM correlations was manifested by the increase of  $1/T_{\rm 1}T\chi$ below 100~K, by the presence of broad magnetic peaks observed below 50~K in the ND measurements, and by the broad peak in $\chi$($T$) at $\approx$ 15~K\@. 

    There are two factors which govern the magnetic properties of this system, $viz.$ frustration and site inversion. 
    We successfully prepared a polycrystalline sample of ${\rm CoAl}_2{\rm O}_4$ with a low inversion parameter $x$ = 0.057(20). 
   Site disorder creates increased randomness of exchange interactions in the already frustrated compound.\cite{Hanashima2013} 
   The suppression of the value of magnetic entropy at high temperatures from the value expected for $S$ = 3/2 suggests that the effective spin of the Co at the $A$-site is smaller than 3/2, which could be a manifestation of both spin frustration and Co-Al site inversion effects. 
    The presence of dynamic short-range magnetic ordering observed in neutron scattering experiments  below $\sim$ 50~K can be  caused mainly by the site inversion effects.
     As pointed out previously, although a spin-liquid ground state has been proposed  from magnetic susceptibility measurements for ${\rm CoAl}_2{\rm O}_4$ with low values of $x$,\cite{Suzuki2007, Hanashima2013} our NMR and ND measurements conclusively proved the existence of a long-range AFM-ordered state below 10~K\@. 
    Since site inversion plays a crucial role in determining the magnetic properties of this spinel compound,  it will also be interesting to perform systematic NMR and ND measurements on  highly disordered (high $x$) ${\rm CoAl}_2{\rm O}_4$ samples to investigate how the system changes from the AFM state to a spin glass state.

\begin{acknowledgments}
    We thank Dr.~Vladimir Tsurkan for providing us the heat capacity data of the nonmagnetic reference compound $\rm {ZnAl_2O_4}$ in Ref. \onlinecite{Tristan2008}. 
   This research was supported by the U.S. Department of Energy, Office of Basic Energy Sciences, Division of Materials Sciences and Engineering. Ames Laboratory is operated for the U.S. Department of Energy by Iowa State University under Contract No.~DE-AC02-07CH11358.
\end{acknowledgments}


\begin{thebibliography}{9}
\bibitem{Roth1964a} W. L. Roth, J. Phys. (Paris) {\bf 25}, 507 (1964).
\bibitem{Bergman2007} D. Bergman, J. Alicea, E. Gull, S. Trebst, and L. Balents, Nature Phys. {\bf 3}, 487 (2007).
\bibitem{Zaharko2011} O. Zaharko, N. B. Christensen, A. Cervellino, V. Tsurkan, A. Maljuk, U. Stuhr, C. Niedermayer, F. Yokaichiya, D. N. Argyriou, M. Boehm, and A. Loidl, Phys. Rev. B {\bf 84}, 094403 (2011).
\bibitem{Tristan2005} N. Tristan, J. Hemberger, A. Krimmel, H.-A. Krug von Nidda, V. Tsurkan, and A. Loidl, Phys. Rev. B {\bf 72}, 174404 (2005).
\bibitem{Ramirez1994} A. P. Ramirez, Annu. Rev. Mater. Sci. {\bf 24}, 453 (1994). 
\bibitem{Suzuki2007} T. Suzuki, H. Nagai, M. Nohara, and H. Takagi, J. Phys.: Condens. Matter {\bf 19}, 145265 (2007).
\bibitem{Zaharko2010} O. Zaharko,  A. Cervellino, V. Tsurkan, N. B. Christensen, and A. Loidl, Phys. Rev. B  {\bf 81}, 064416 (2010).
\bibitem{MacDougall2011} G. J. MacDougall, D. Gout, J. L. Zarestky, G. Ehlers, A. Podlesnyak, M. A. McGuire, D. Mandrus, and S. E. Nagler, Proceedings of the National Academy of Sciences {\bf 38}, 15693 (2011).
\bibitem{Tristan2008} N. Tristan, V. Zestrea, G. Behr, R. Klingeler, B. B\"uchner, H.-A. Krug von Nidda, A. Loidl, and V. Tsurkan, Phys. Rev. B {\bf 77}, 094412 (2008).
\bibitem{Hanashima2013} K. Hanashima, Y. Kodama, D. Akahoshi, C. Kanadani, and T. Saito, J. Phys. Soc. Jpn. {\bf 82}, 024702 (2013).
\bibitem{Carvajal1993} J. Rodr\'iguez-Carvajal, Physica B {\bf 192}, 55 (1993); see also {\tt www.illeu/sites/fullprof/}.
\bibitem{Maljuk2009} A. Maljuk, V. Tsurkan, V. Zestrea, O. Zaharko, A. Cerellino, A. Loidl, and D. N. Argyriou, J. Cryst. Growth  {\bf 311}, 3997 (2009). 
\bibitem{Hagiwara2010} M. Hagiwara, S. Kimura, N. Nishihagi, T. Suzuki, M. Nohara, H. Takagi, and K. Kindo, J. Low Temp. Phys. {\bf 159}, 11 (2010). 
\bibitem{Meschede1980} D. Meschede, F. Steglich, W. Felsch, H. Maletta, and W. Zinn, Phys. Rev. Lett. {\bf 44}, 102 (1980).
\bibitem{Walker1977} L. R. Walker and R. E. Walstedt, Phys. Rev. Lett.  {\bf 38}, 514 (1977).
\bibitem{Miyatani1965} K. Miyatani, K. Kohn, S. Iida, and H. Kamimura, J. Phys. Soc. Jpn. {\bf 20}, 471 (1965).
\bibitem{de -Jongh} {\it Magnetic Properties of Layered Transition Metal Compounds} edited by L. J. de Jongh (Klewer, Dordrecht, 1989). 
\bibitem{Fukai1996} T. Fukai, Y. Furukawa, S. Wada, and K. Miyatani, J. Phys. Soc. Jpn. {\bf 65}, 4067 (1996).
\bibitem{Tsuda1968} T. Tsuda, A. Hirai, and H. Abe, Phys. Lett. {\bf 26A}, 463 (1968). 
\bibitem{Furukawa1999} Y. Furukawa, S. Wada, T. Kajitani, and S. Hosoya, J. Phys. Soc. Jpn. {\bf 68}, 346 (1999).
\bibitem{Abragam1955} A. Abragam, J. Horowitz, and M. H. L. Pryce, Proc. Roy. Soc. (London), Ser. A {\bf 230}, 169 (1955).

\bibitem{Narath_Hyper} A. Narath, {\it Hyperfine Interactions} edited by A. J. Freeman and R. B. Frankel (Academic Press, New York, 1967). 
\bibitem{Freeman1965}A. J. Freeman and R. E. Watson, {\it Magnetism} edited by G. T. Rado and H. Shul (Academic Press, New York, 1965) Vol. 2.
\bibitem{Narath1967} A. Narath, Phys. Rev. {\bf 162}, 320 (1967). 
\bibitem{Moriya1963} T. Moriya, J. Phys. Soc. Jpn. {\bf 18}, 516 (1963).
\bibitem{Mahajan1998} A. V. Mahajan, R. Sala, E. Lee, F. Borsa, S. Kondo, and D. C. Johnston, Phys. Rev. B {\bf 57}, 8890 (1998).
\bibitem{Roth1964b} W. L. Roth, J. Phys. Chem. Solids {\bf 25}, 1 (1964)
\bibitem{Krimmel2006a} A. Krimmel, V. Tsurkan, D. Sheptyakov, and A. Loidl, Physica B {\bf 378-380}, 583 (2006).
\bibitem{Nakatsuka2003} A. Nakatsuka Y. Ikeda, Y. Yamasaki, N. Nakayama and T. Mizota,  Solid State Commun. {\bf 128}, 85 (2003).

\end{thebibliography}
\end{document}